\documentclass{elsart}
\usepackage{graphicx,natbib,amssymb}

\newcommand{\wcap}{0.476\textwidth}

\begin{document}

\begin{frontmatter}

\title{Causes and Consequences of Magnetic Field Changes in Neutron Stars}

\author{M. Ruderman}\ead{mar@astro.columbia.edu}  

\address{Department of Physics and Columbia Astrophysics Laboratory
\newline Columbia University, New York, NY}

\begin{abstract}
Because of the quantum fluid properties of a neutron star core's neutrons
and protons, its magnetic field is expected to be coupled strongly to its
spin.  This predicts a simple evolution of the surface-field of such stars
as they spin down or, less commonly, are spun up.  Consequences and comparisons with
observations are given for properties of solitary spinning down pulsars, 
including their glitches and spin-down ages, X-ray pulsars, and the formation and 
pulse characteristics of Millisecond Pulsars.
For none of these is there a present conflict between model predictions and 
what has been observed.
\end{abstract}


\end{frontmatter}

\section{Introduction}

Enough is understood about the dynamics of the components of a standard 
(non-magnetar, non-strange) neutron star (NS) to support what should be a 
reliable description of what happens within a spinning magnetized NS as it ages
and spins down or, in rarer cases, when it is spun up.

In a cool core below the crust of a spinning NS superconducting protons coexist
with more abundant superfluid neutrons (SF-n) to form a giant atomic nucleus
which contains within it a neutralizing sea of relativistic degenerate 
electrons. The neutrons rotate with a spin-period $P$ (sec) 
$\equiv 2\pi /\Omega$ only by forming a
nearly uniform array of corotating quantized vortex lines parallel to the spin
axis, with an area density $n_v \sim 10^4~{\rm cm}^{-2}~P^{-1}$.
The array must contract (expand) when the NS spins up (down). 
In stellar core neutron spin-up or spin-down, a vortex a distance $r_\perp$ 
from the spin-axis generally moves outward with a velocity $v_v = r_\perp (\dot{P}
/ 2 P)$ until $r_\perp$ reaches the core's neutron superfluid radius ($R$).
Any stellar magnetic field passing below the stellar crust must, in order to 
penetrate through the core's superconducting protons (SC-p), become a very dense
array of quantized flux-tubes ($n_\Phi \sim 5 \times 10^{18} ~ B_{12} ~ 
\rm{cm}^{-2}$ with $B$ the local average magnetic field).
Each tube carries a flux $2 \times 10 ^{-7} ~ \rm{G} ~ \rm{cm}^2$ and a magnetic field
$B_c \sim 10^{15}~\rm{G}$.\footnote{
This assumes Type II proton superconductivity in the NS core, below the crust,
the common result of many calculations.  If it were Type I, with many thin
regions of $B > $ several $B_c$, and $B \sim 0$ in between\cite{ref10}, the 
impact on surface $B$ of changing NS spin proposed below would not change
significantly. If, however, $\langle B \rangle$, the locally averaged 
$B$ inside the NS core exceeds
a critical field somewhat greater than $B_c$, the core's protons would not
become superconducting.
This may well be the case for most (or all) ``magnetars''\cite{ref11}.
}
\begin{figure*}
\centerline{
\includegraphics*[width=4in]{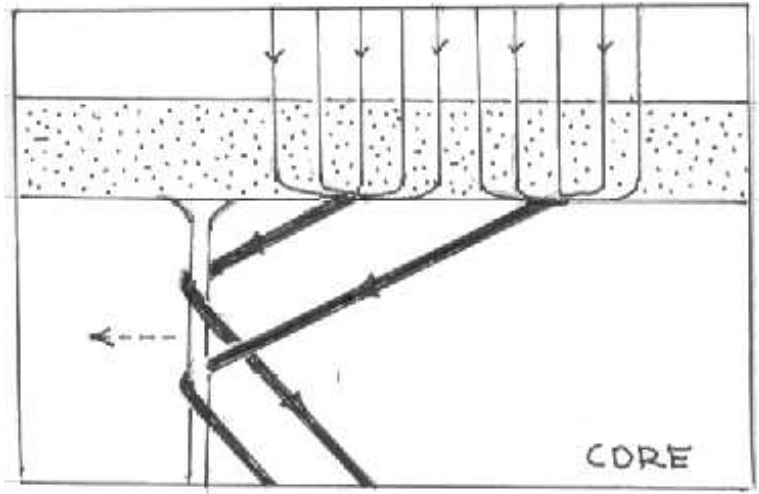}}
\caption{A moving quantized vortex-line in a NS core's superfluid neutrons
pulling a pair of the core's proton superfluid quantized flux-tubes anchored
in the star's solid, conducting crust (shown dotted).}
\label{f1}
\end{figure*}
The initial magnetic field within the core of a neutron star is expected to have
both toroidal and very non-uniform poloidal components.
The web of flux-tubes formed after the transition to superconductivity is then
much more complicated and irregular than the neutron vortex-array as well as of
order $10^{14}$ times more dense.
Because of the velocity dependence of the short range nuclear force between 
neutrons and protons, there is a strong interaction between the 
neutron-superfluid's vortex-lines and the proton-superconductor's flux-tubes
if they come closer to each other than about $10^{-11} \rm{cm}$.
Consequently, when $\dot{P} \neq 0$ flux tubes will be pushed (or pulled)
by the moving neutron vortices\cite{ref1,ref2,ref3,ref4,ref5,
ref6,ref7,ref8,ref9}.  
A realistic flux-tube array will be forced to move along with a changing
SF-n vortex array which threads it as long as the force at a vortex-line
flux-tube juncture does not grow so large that vortex lines cut through
flux-tubes.
The drag on moving flux-tube arrays from their small average velocities 
($\dot{r}_\perp < 10^{-5} \rm{cm~s}^{-1}$) in spinning-down pulsars, cool (old)
enough to have SF-n cores, seems far too small to cause such 
cut-through.\footnote{
Jones\cite{ref30} has recently found that electron scattering on flux-tube cores
allows easier passage of flux-tubes through the SC-p than had been previously 
estimated (e.g. \cite{ref5}).  
In addition, an expected motion-induced flux-tube
bunching instability would allow easy co-motion of flux-tubes with the local
electron plus SC-p fluid in which they are embedded \cite{ref12}.
}

The 
main quantitative uncertainty in the model described below is the maximum 
sustainable shear-strain ($\theta_m \sim 10^{-4}-10^{-3}$ ?) on the conducting
crust, which anchors core flux-tubes (cf. Fig 1), before the crust yield-strength
 is exceeded.
An estimate\cite{ref8} for that maximum sustainable crustal shear-stress, compared
to that from the $\langle B^2 \rangle/8\pi \sim \langle B \rangle B_c/8\pi$ of 
the core's flux-tube array,
supports a NS model in which the crust yields before the core's flux-tubes are 
cut through by its moving SF-n vortex array,  as long as $B_{12} \gtrsim 1 $.
Even for much smaller $B_{12}$, flux-tube anchoring by the conducting crust 
would result in such cut-through only when the NS's spin-down age ($P/2\dot{P}$)
exceeds the crust's Eddy current dissipation time ($\sim 10^7$ yrs.).
Then in most observationally relevant regimes the motion of the magnetic 
flux-tube array near the top of the NS core (and $B$ at the crust surface above it) follows that of the SF-n vortex array which threads it.
This forms the basis of a very simple model for describing predicted changes
in pulsar magnetic fields during NS spin-up or spin-down which agrees well with
a variety of different families of pulsar observations.

\section{Magnetic field changes in spinning up neutron stars}

NS spin-up, when sustained long enough so that one of the above criteria
for relaxation of shear-stress from crust-anchored magnetic flux 
before cut-through is met,
leads to a ``squeezing'' of surface {\bf B} toward the NS spin-axis.
After a large decrease in spin-period from an initial $P_0$ to $P \ll P_0$
all flux would enter and leave the core's surface from the small area within
a radius $R(P/P_0)^{1/2}$ of the NS's spin-axis.
This surface {\bf B}-field change is represented in Figs 2-3 for the special case
when the magnetic flux which exits the NS surface from its upper (lower) 
spin-hemisphere
returns to the stellar surface in it's lower (upper) one.
Potentially observable features of such a ``spin-squeezed'' surface {\bf B}
configuration include the following.

\begin{figure}[tbh]
  \centerline{
    \includegraphics[height=2.75in]{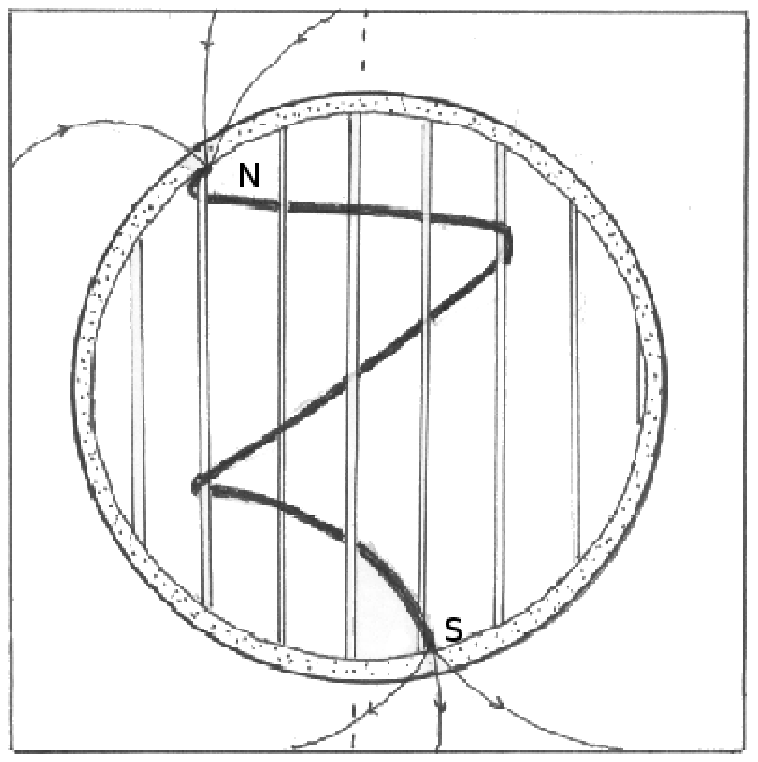}\hfill
     \includegraphics[height=2.75in]{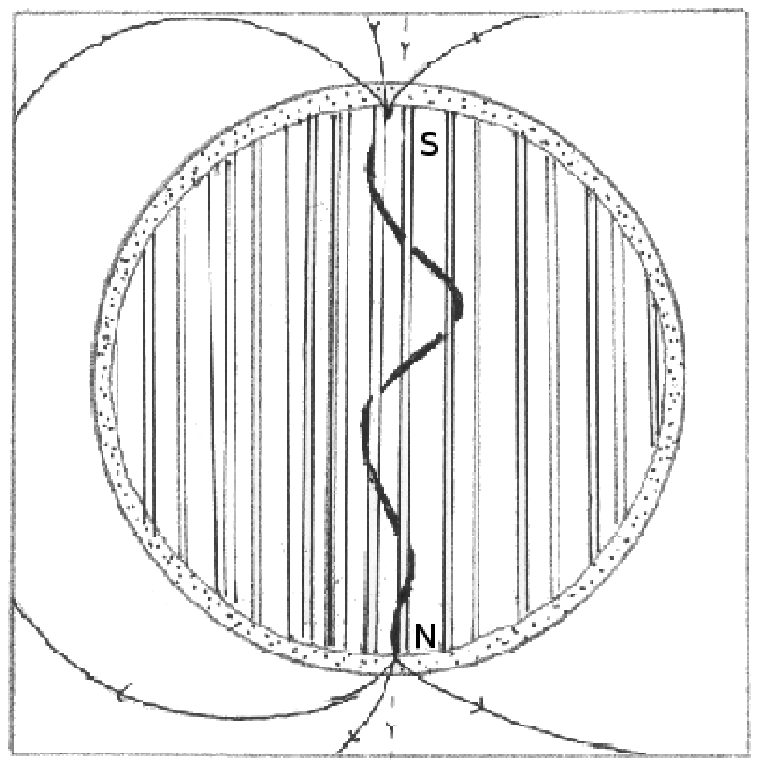}\hfill
   }
  \begin{minipage}{\wcap}
    \caption{A single flux-tube (one of $10^{31}$) and some of the NS's 
	arrayed vortices (8 of $10^{17}$).}
  \end{minipage} \hfill
  \vspace{.6cm}
  \begin{minipage}[tr]{\wcap}
    \caption{The flux-tube and vortex array of Fig. 2 after a large stellar 
	spin-up.}
    \vspace{.6cm}
  \end{minipage}
\end{figure}

\newcounter{Lcount}
\begin{list}{\alph{Lcount})}
  {\usecounter{Lcount}
  \setlength{\rightmargin}{\leftmargin}}
\item A dipole moment nearly aligned along the NS spin-axis.
\vspace*{.6cm}

\item A greatly diminished polar cap spanning the ``open'' field lines when
$P/P_0 \rightarrow 0$.  
For $P < P_0 (\Delta/R)(\Omega_0 R/c)^{1/2}$ with
$\Delta$ the crust thickness ($\sim 10^{-1} R$), the canonical polar cap radius,
$ r_p \equiv R (\Omega R/c)^{1/2} $, shrinks to 
$ r_p' \equiv \Delta (\Omega R/c)^{1/2} $
\vspace*{.6cm}

\item A $B$-field just above the polar cap which has almost no curvature.
\end{list}

If the pre-spin-up surface {\bf B} has a sunspot-like
configuration (i.e. flux returning to the NS surface in the same hemisphere as
that from which it left), the spin-up-squeezed field change is represented in
Figs 4 and 5.
In this case, potentially observable features when $P \ll P_0$ include the 
following.

\begin{figure}[bth]
  \centerline{
    \includegraphics[height=2.75in]{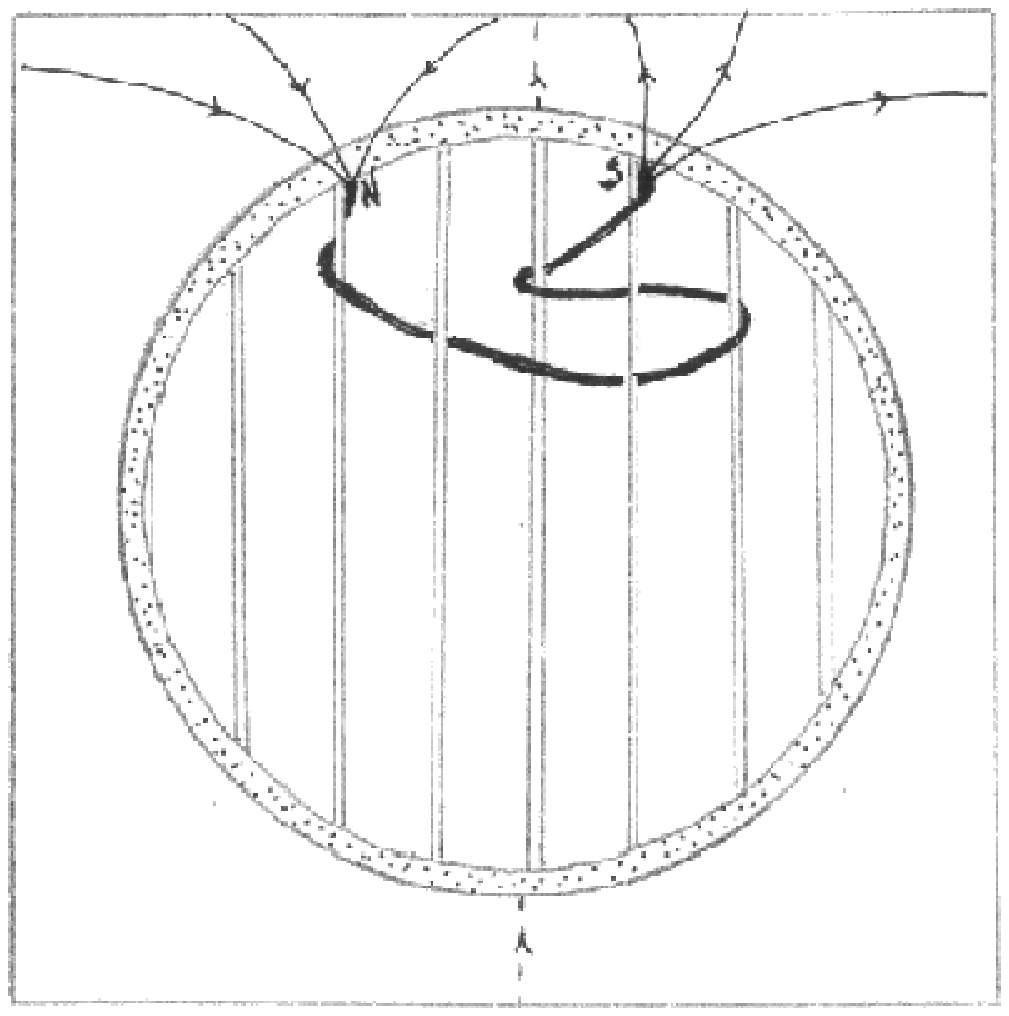}\hfill
    \includegraphics[height=2.75in]{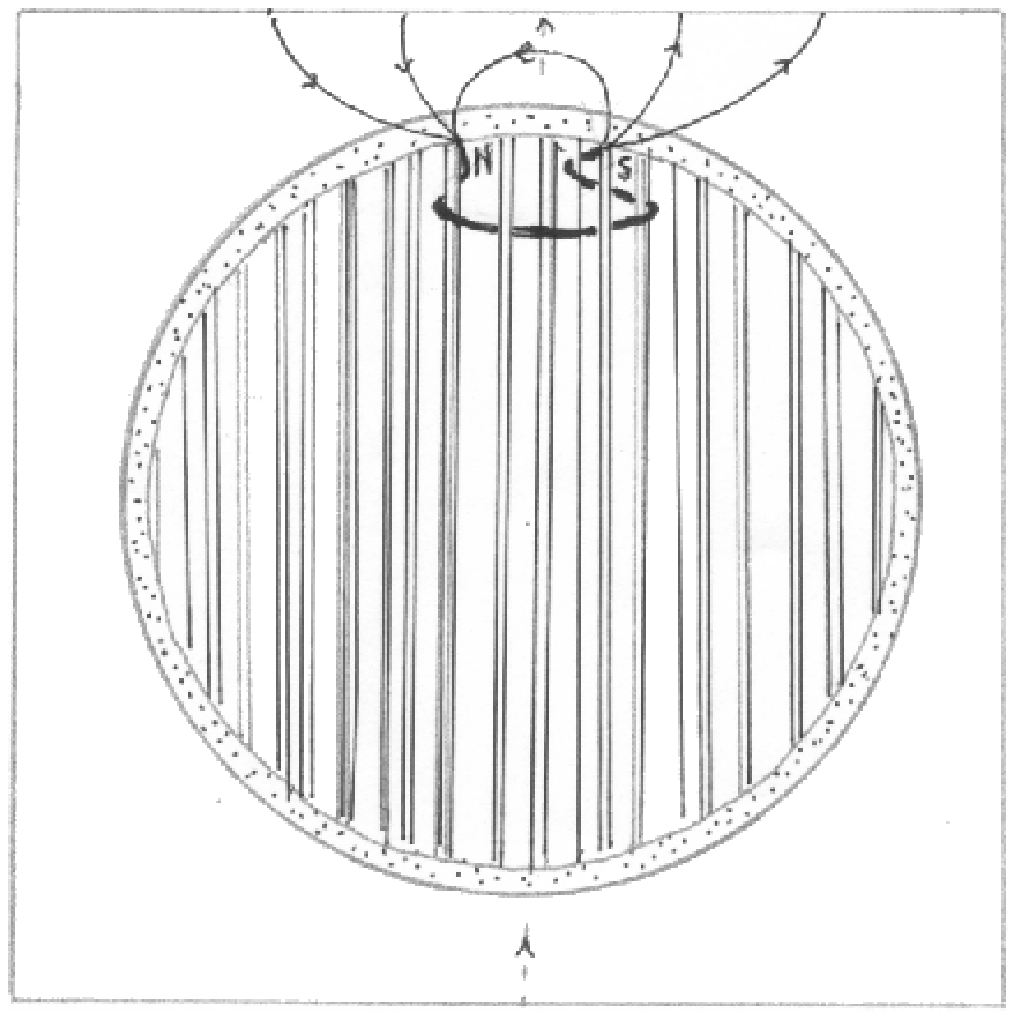}\hfill
   }
  \begin{minipage}{\wcap}
    \caption{A single flux-tube, part of a sunspot-like $B$-field geometry in
	which flux from a spin-hemisphere of the surface returns to the surface
	in that same hemisphere.}
    \vspace{-.60cm}
  \end{minipage} \hfill
  \vspace{.6cm}
  \begin{minipage}[tr]{\wcap}
    \caption{The flux-tube and vortex array of Fig. 4 after a large stellar 
	spin-up.}
    \vspace{.9cm}
  \end{minipage}
\end{figure}

\begin{list}{\alph{Lcount})}
  {\usecounter{Lcount}
   \setcounter{Lcount}{3}
  \setlength{\rightmargin}{\leftmargin}}
\item A pulsar dipole moment nearly orthogonal to the NS spin-axis, and
\vspace*{.6cm}

\item positioned at the crust-core interface.
\vspace*{.6cm}

\item A dipole moment ({\boldmath$\mu$}), or more precisely the component 
of {\boldmath$\mu$} perpendicular to {\boldmath$\Omega$}, reduced from its 
pre-spin-up size:
\begin{equation}\label{eq1}
	{{\mu_\perp(P)}\over{\mu_\perp(P_0)}} 
	\sim \left({P \over P_0}\right)^{1/2}.
\end{equation}
\end{list}

A more general (and very probably more realistic) pre-spin-up configuration 
has flux emitted from one spin-hemisphere returning to the stellar 
surface in both, as in Fig. 6. 
Spin-up squeezing then typically gives the surface field configuration represented in 
Fig. 7, a spin-squeezed, nearly orthogonal dipole on the NS spin-axis
with properties (d), (e), and (f),
together with an aligned dipole on the spin-axis whose external field is 
well-represented by North and South poles a distance $2(R-\Delta)$ apart.
Further spin-up could lead to the Figs. 8 and 3 configuration; 
that of Fig. 9 and 5
would be realized only if $S_2$ of Figs. 6 and 7 is negligible.

\begin{figure}[tbh]
  \centerline{
    \includegraphics[height=2.75in]{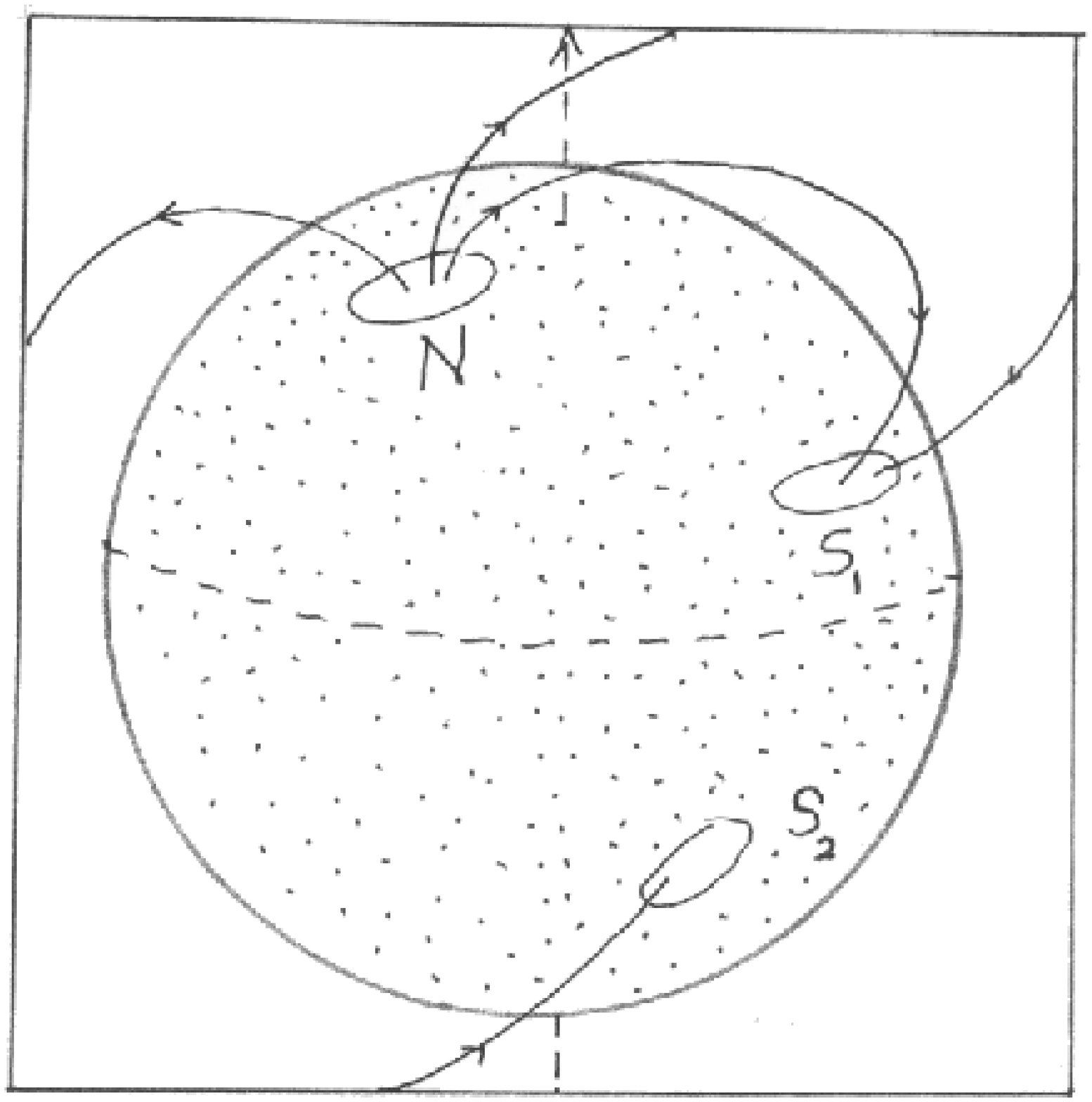}\hfill
    \includegraphics[height=2.75in]{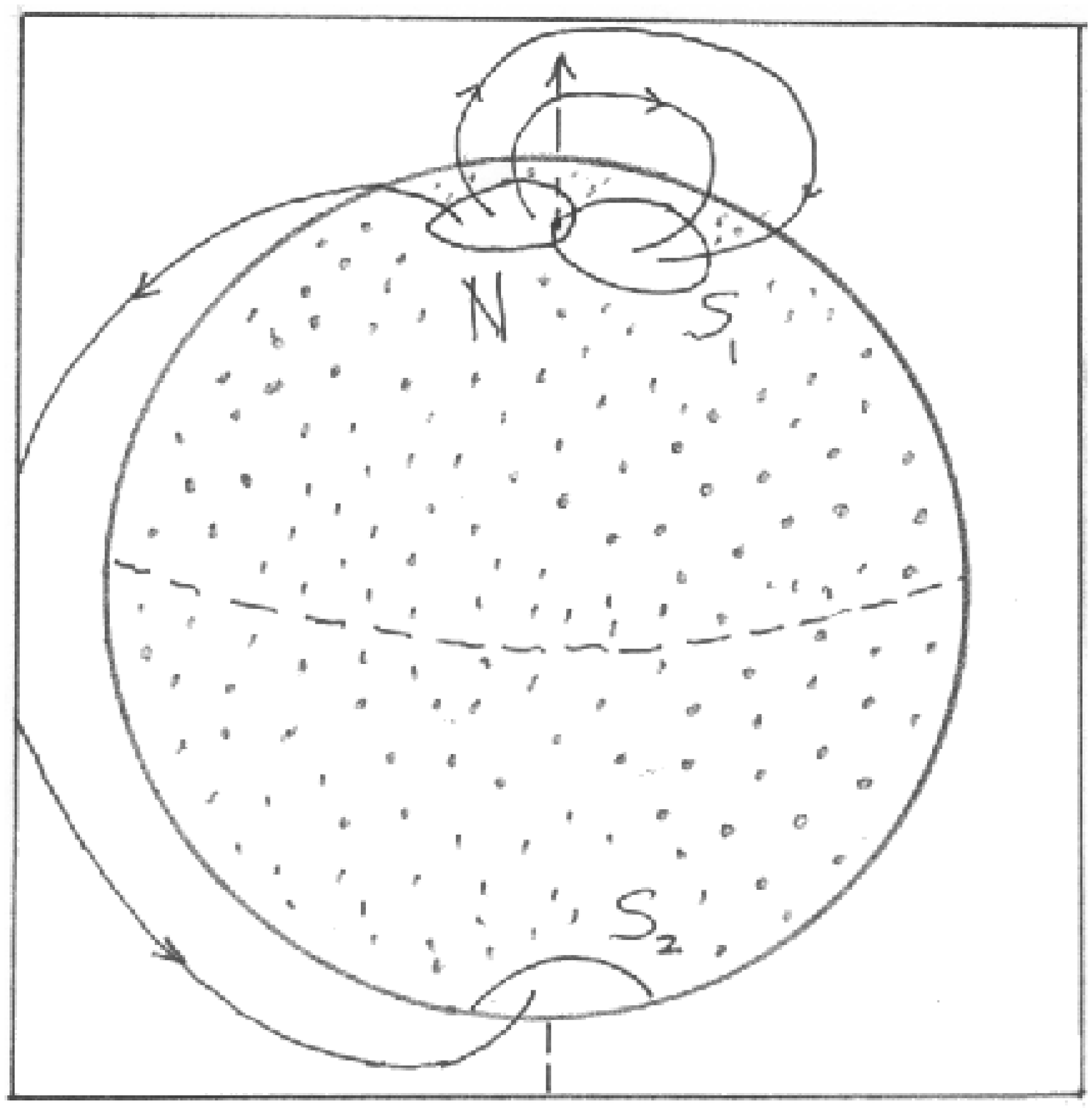}\hfill
   }
  \begin{minipage}{\wcap}
    \caption{A surface field which has flux of both Fig. 2 and Fig. 4
	configurations. }
    \vspace{.6cm}
  \end{minipage} \hfill
  \vspace{.6cm}
  \begin{minipage}[tr]{\wcap}
    \caption{The field from Fig. 6 after a large stellar spin-up.}
    \vspace{.6cm}
  \end{minipage}
\end{figure}

\begin{figure}[tbh]
  \centerline{
    \includegraphics[height=2.75in]{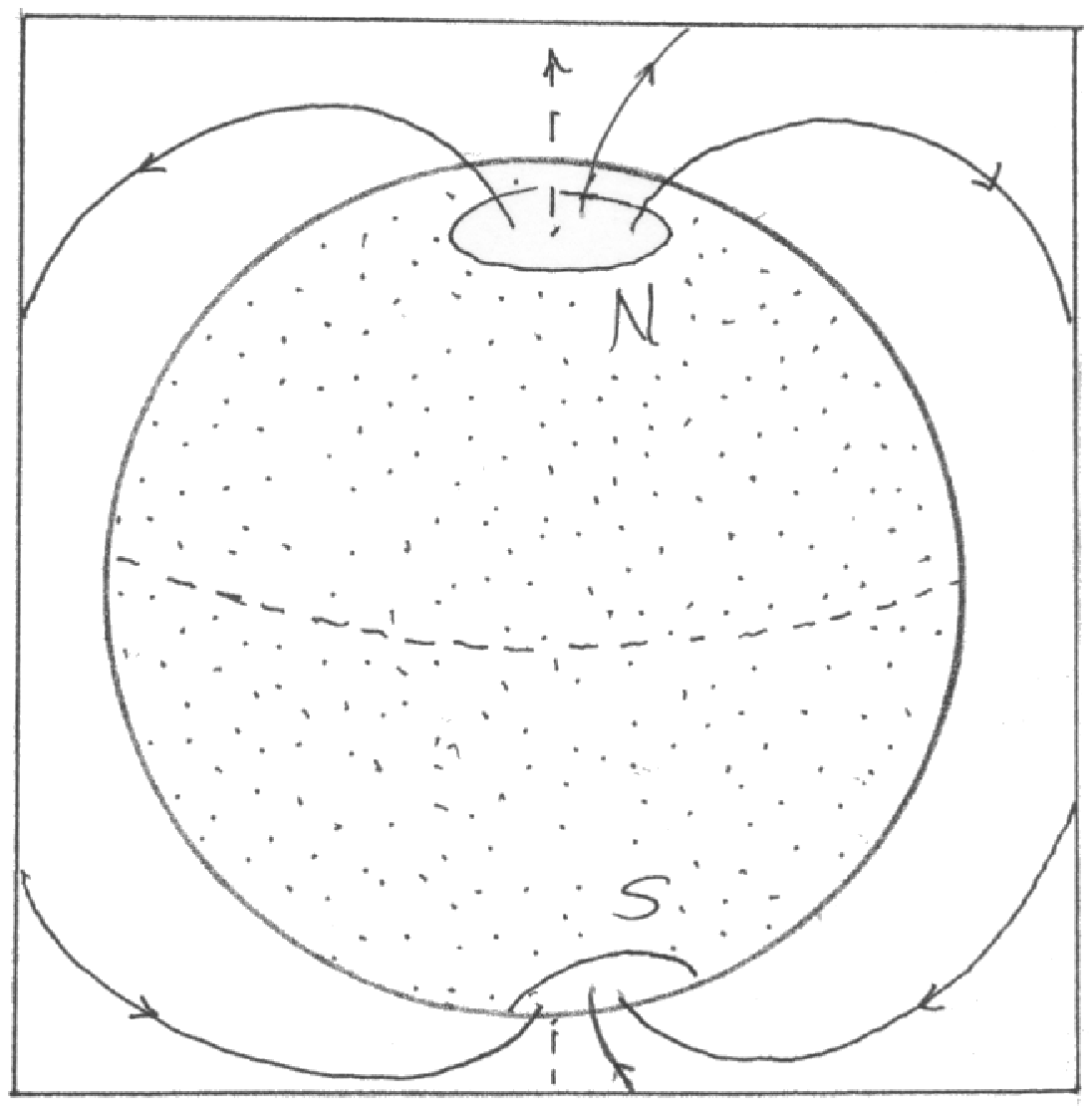}\hfill
    \includegraphics[height=2.75in]{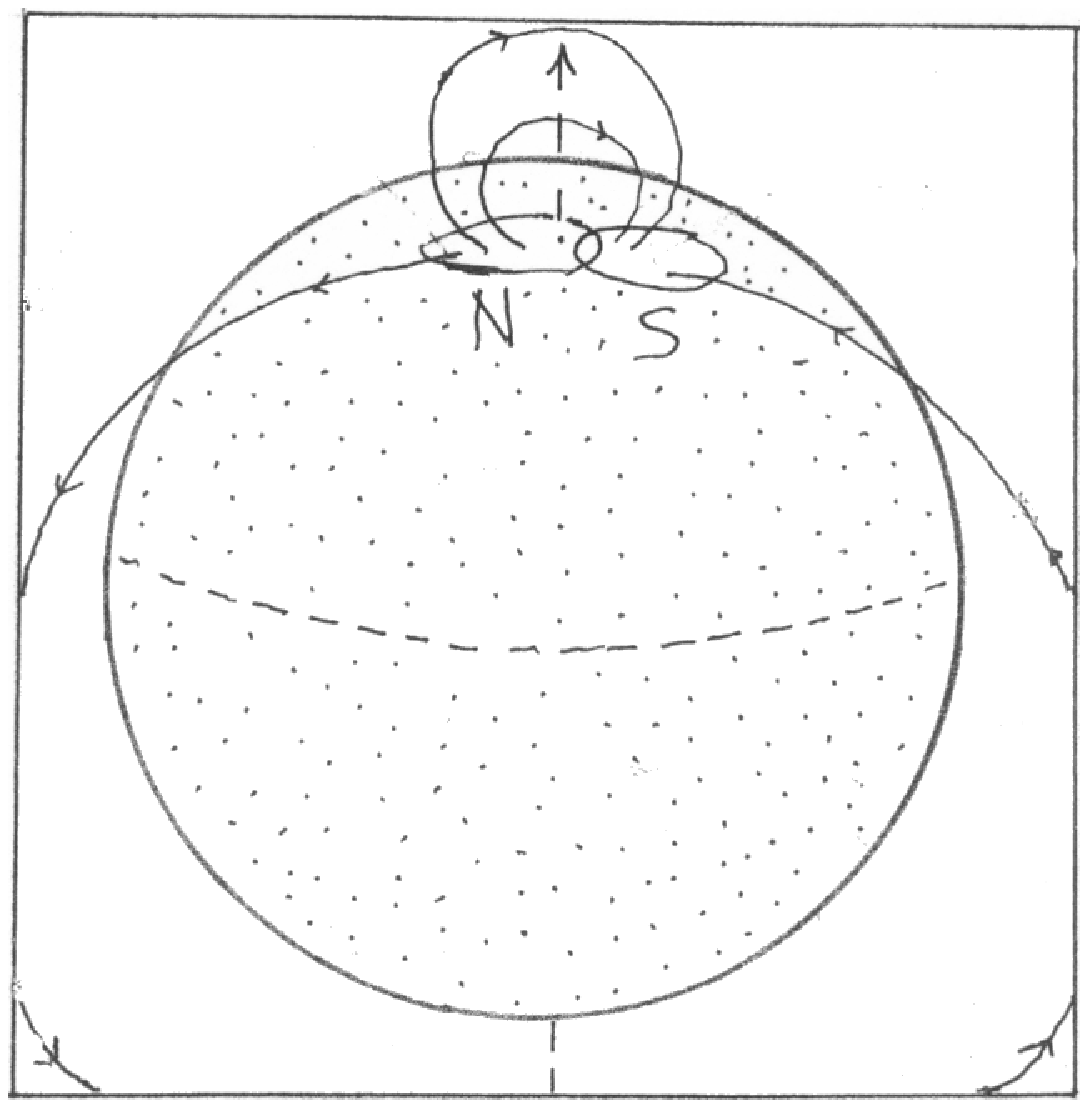}\hfill
   }
  \begin{minipage}{\wcap}
    \caption{The field from Fig. 7 after further spin-up.}
    \vspace{1cm}
  \end{minipage} \hfill
  \vspace{.6cm}
  \begin{minipage}[tr]{\wcap}
    \caption{The field from Fig. 6 after large spin-up when the $S_2$
	contribution to Fig. 7 is negligible.}
    \vspace{.6cm}
  \end{minipage}
\end{figure}


\section{Magnetic field changes in spinning down neutron stars}

Consequences of the coupling between a spin-down expansion of a NS's SF-n 
vortex-array and its SC-p flux-tubes should appear in several observable phases
which begin after the NS has cooled enough that the vortex-line array and
the flux-tube one have both been formed (typically after about $10^3$ yrs.).


\begin{figure}[tbh]
  \centerline{
    \includegraphics[height=2.75in]{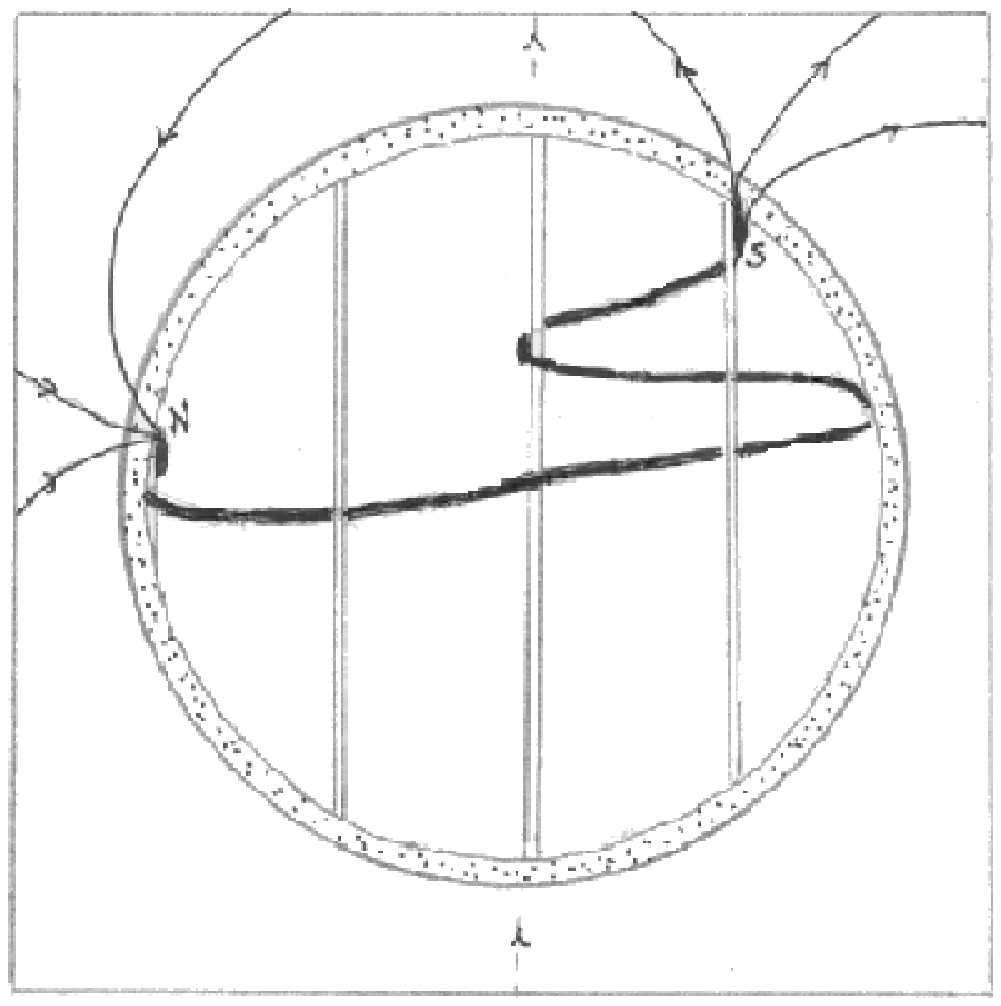}\hfill
    \includegraphics[height=2.75in]{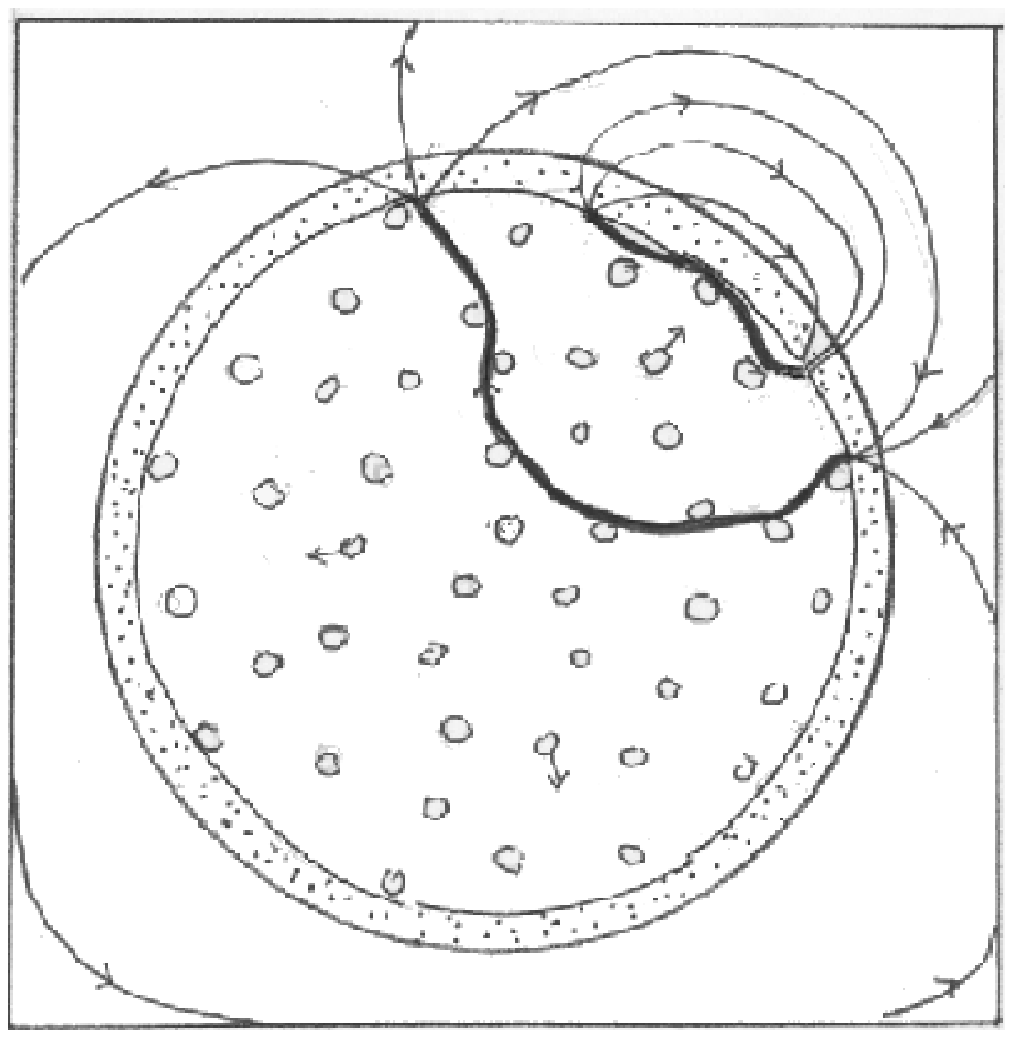}\hfill
   }
  \begin{minipage}{\wcap}
    \caption{The flux-tube and vortex-array of Fig. 4 after some spin-down.
	The expanded configuration would not differ in an important way if
	it had begun as that of Fig. 2.}
    \vspace{.6cm}
  \end{minipage} \hfill
  \vspace{.6cm}
  \begin{minipage}[tr]{\wcap}
    \caption{The equatorial plane, viewed from above, of a configuration like
	that of Fig. 10, but with two flux-tubes.  One tube is being expelled
	into the crust by the expanded vortex array and will ultimately be 
	eliminated by reconnection.}
  \end{minipage}
\end{figure}


\begin{list}{\alph{Lcount})}
  {\usecounter{Lcount}
   \setcounter{Lcount}{0}
  \setlength{\rightmargin}{\leftmargin}}

\item As in Eqn (\ref{eq1}), except that $P > P_0$, $\mu_\perp(P)$
initially grows as $P^{1/2}$.  This increase is initially not 
sensitive to the configuration of surface {\bf B} (cf. Fig. 10).
\vspace*{.6cm}

\item  When $P \sim \rm{several}~ P_0$, a good fraction of a NS's core
flux-tubes will have been pushed outwards from the spin-axis to $r_\perp
\sim R$.  These cannot, of course, continue to move outward (Fig. 11) so
that Eqn (\ref{eq1}) no longer holds.
Rather, the mixture of expanding and crust-constrained flux-tubes gives:
\begin{equation}\label{eq2}
	{{\mu_\perp(P)}\over{\mu_\perp(P_0)}} 
	\sim \left({P \over P_0}\right)^{\hat{n}} 
	~~~~~~~~~~~~~~~~ (0 < \hat{n} < 1/2)
\end{equation}
with the exact value of $\hat{n}$ dependent on details of a core's $B$-field
configuration.
\vspace*{.6cm}

\item The crust can delay, but not indefinitely prevent, expulsion of this 
magnetic field from the NS.
When $P \sim \rm{several}~ P_0$, intertwined vortex plus flux
which have been pushed into the core-crust interface will stress the crust
enough to exceed its shear-strength (Sect. 4 and Figs. 15 and 16).  
Then crust movements begin that lead to $B$-field reconnections.  
Flux that is threaded 
by SF-n vortex lines that have not yet reached $r_\perp \sim R$, and thus have
not yet disappeared in this way, are the remaining source for the NS's 
dipole moment.  
The sum of all this remaining flux $\propto$ 
the total number of remaining vortex-lines ($\propto \Omega$).  
Then, Eqn (\ref{eq2}) holds with 
$\hat{n}=-1$.
\vspace*{.6cm}


\item When this remaining $B$ at the crust bottom $(\propto \Omega)$ drops 
to and below $\sim 10^{12} ~\rm{G}$, shear-stress on the crust would no longer 
be expected to
exceed the crust's yield-strength.  The NS's surface $B$ may then lag that at
the base of its crust by as much as $10^7$ yrs., the crust's Eddy current 
dissipation time.
\end{list}


\section{Comparisons of pulsar expectations with model expectations}

Fig. 12 shows observationally inferred surface dipole fields ($B$) as a function
of their $P$
for about $10^3$ radiopulsars ($B$ is calculated from measured $P$ and 
$\hat{P}$, ${\rm I}\dot{\Omega} = - \mu_{\perp}^2~\Omega^3~c^{-3}$; 
$B=\mu_{\perp} R^{-3}$ and ${\rm I} = 10^{45} \rm{g}~ \rm{cm}^2$.).  Segments of 
$B(P)$ based upon the model of Sects 2 and 3, are shown for a typical 
pulsar by the doubled or single solid lines.

\begin{enumerate}

\item Point $A$ is the $(B,P)$ where, typically, flux-tubes and vortex lines 
begin coexistence.
\vspace*{.6cm}

\item $(A \rightarrow C)$ is the expanding array prediction of Sect. 3(b): 
$B \propto P^{\hat{n}}$ with the model prediction $0 < \hat{n} < 0.5$.  
The index $\hat{n}$ is known only in the several cases where $\ddot{P}$ is also 
measured: $\hat{n} = +0.3, 0.1, 0.6, 0.1 $ \cite{ref13,ref14,ref15,ref16}.
\vspace*{.6cm}

\begin{figure*}
\centerline{
\includegraphics*[width=5.5in]{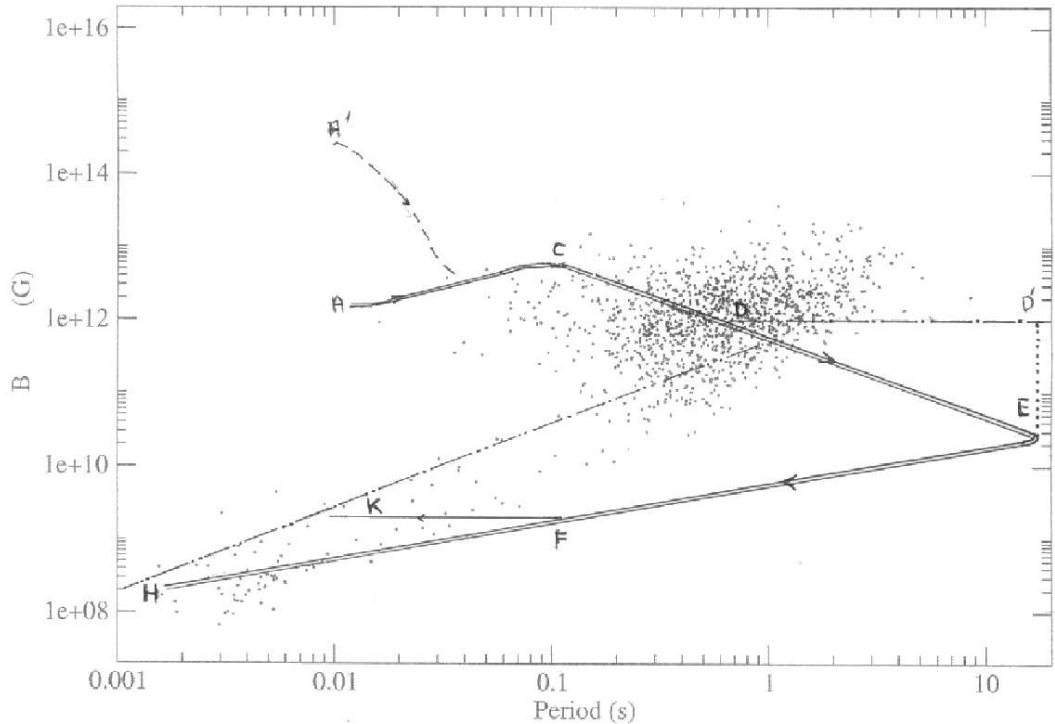}}
\caption{Dipole-$B$ observed on pulsar surfaces (inferred from measured
{\em P} and {\em \.P}) as a function of pulsar period ({\em P}) \cite{ref30}.
The solid line segments are the evolutionary segments for $B$ of Sect. 4,
based upon the model of Sects. 2 and 3.  The dash-dot diagonal is the 
steady state spin-up line from an accretion disk.  The horizontal
$(D \rightarrow D')$ is that for a NS surface above a core surface
$(D \rightarrow E)$.  
}
\label{f13}
\end{figure*}


\item  $(C \rightarrow D)$ is the flux-expulsion and continual reconnection 
segment of Sect. 3(e). The model predicts 
$\langle \hat{n}\rangle = -1$ for $\hat{n}$ averaged over the 
$(C \rightarrow D)$ history of any one pulsar.  
Reliable $\ddot{P}$ are not generally 
measurable in this $(B,P)$ region.  However comparison of the spin-down times
$P/2\dot{P}$ with actual pulsar ages, inferred from probable distance traveled 
since birth\cite{ref17}, give $\langle \hat{n}\rangle = -0.8 \pm 0.4$, not 
inconsistent with the model prediction.
\vspace*{.6cm}

\item  $(D \rightarrow E)$ is the core-surface/crust-base $B$ evolution
for $\sim 10^{10}$ yrs.  The horizontal $(D \rightarrow D')$ is the NS crust's
surface field, remaining near $10^{12}$ G for $\sim 10^7$ yrs. as discussed
in Sect. 3(d).  
This segment should be characteristic of typical ``X-ray pulsars''
(NSs in binaries spun up or down by active companions through a wide range of 
$P$ (e.g. Hercules X-1 with $P \sim 1$s to Vela X-1 
with $P \sim 10^3$s) until
crustal Eddy current decay allows a $(D' \rightarrow E)$ decay from some 
$D'$ region.
\vspace*{.4cm}

A small minority of NSs, after $(D' \rightarrow E)$ segments, will be resurrected 
by accretion from a previously passive White Dwarf companion which now overflows its
Roche lobe (LMXBs).  
These NSs have entered into the spin-up phase of Sect. 2 until they reach a steady
state on the canonical ``spin-up line'' represented by the dot-dashed diagonal
of Fig. 12 (for $\dot{M} = 10^{-1} \dot{M}_{\small Eddington}$).
\vspace*{.6cm}

\item $(E \rightarrow F \rightarrow H)$ is the spin-up segment when the NS 
surface $B$ has the sunspot geometry of Figs. 4, 5, and 9, which allows spin-up
to minimal $P$ before spin-up equilibrium is reached.   
Observations of maximally spun-up millisecond pulsars (MSPs) support the Sect.
2 model for such MSP formation: Sect. 2(d)'s high fraction of MSPs with two 
subpulses $180^\circ$ apart, characteristic of orthogonal 
rotators\cite{ref17,ref18,ref19,ref20,ref21}; 
Sect. 2(e)'s $B$-field geometry, from linear polarization and its frequency 
dependence in such subpulses\cite{ref20}.
\vspace*{.6cm}

\item $(E \rightarrow F \rightarrow K)$ is the track of surface $B$ (here the
total dipole field) predicted after large spin-up from $(E)$ with Fig. 6 
geometry to $(F)$ with Fig. 7 geometry. Further spin-up diminishes only the 
orthogonal component of {\boldmath$\mu$} until an almost aligned rotator 
(Figs. 3 and 8) results when $(K)$ is reached.
X-ray emission from the almost aligned MSP PSR 0437 ($P = 6$ ms) supports a 
(predicted) tiny polar cap area about $(\Delta/R)^2 \sim 10^{-2}$ that
from a central dipole moment configuration for the same $P$ (Sect.
2(b) and refs \cite{ref20,ref18,ref21}).  

\vspace*{.4cm}
Expected consequences for pulsar diplole-{\bf B} changing according to the 
Sects. 2-3 model and Fig. 12 are supported by many kinds of observations.
However, for almost all there is usually another popular explanation 
(e.g. $B$ getting from $(D)$ to $(H)$ just by burial of {\bf B} by accreted 
matter from a companion\cite{ref22,ref23,ref24}).
\end{enumerate}

\section{Pulsar spin-period glitches from spin-induced $B$-field changes}

Moving core flux-tubes continually build up shearing stress in the conducting
crust which anchors $B$-field that traverses it.  If this stess grows to exceed
the crust's yield strength, subsequent relaxation may, at least partly, be 
through relatively sudden crustal readjustments (``crust-breaking'').  
Such events would cause very small spin-up jumps in spinning-down NSs
(spin-period ``glitches'').  The Sect. 2-3 model for the evolution of a core's
flux-tube array suggests glitch details in pulsars similar to those of the two
observed glitch families: Crab-like glitches (C) and the very much larger
giant Vela-like ones (V) of Fig. 13.

\begin{figure*}
\centerline{
\includegraphics*[width=5.5in]{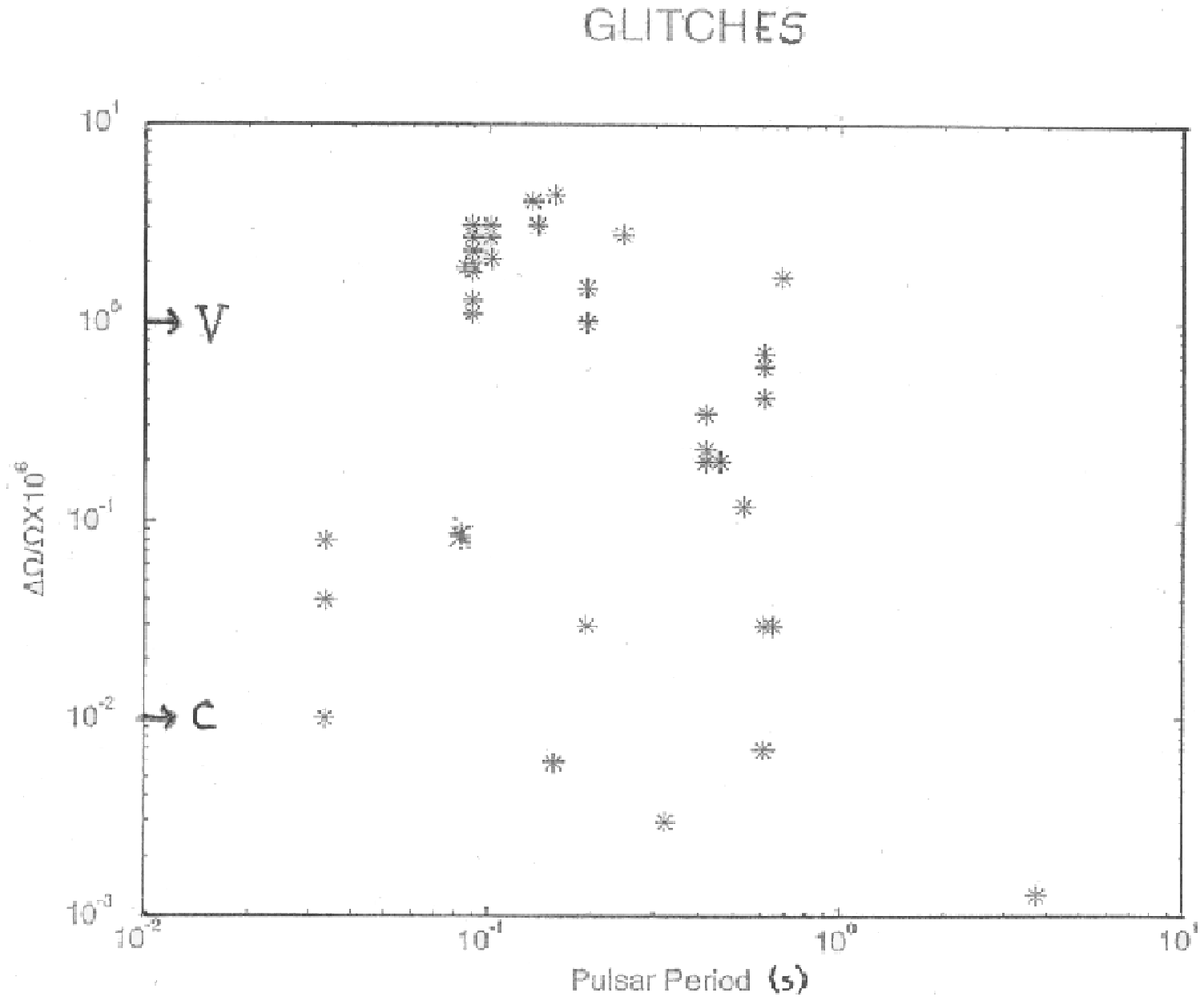}}
\caption{Observed jumps (``glitches'') in pulsar spin-rates 
	$(\Delta \Omega / \Omega)$ of pulsars with various periods ($P$).
	The Vela-like family ({\bf V}) has $\Delta \Omega / \Omega \sim 10 ^{-6}$.
	The Crab-like one ({\bf C}) has $\Delta \Omega / \Omega \sim 10^{-7} -
	10^{-8}$ \cite{ref25,ref26,ref27,ref28,ref34}.}
\label{f14}
\end{figure*}

\begin{list}{\alph{Lcount})}
  {\usecounter{Lcount}
   \setcounter{Lcount}{0}
   \setlength{\rightmargin}{\leftmargin}}

\item {\emph{Crab-like glitches}} In both the $(A \rightarrow C)$ and 
$(C \rightarrow D)$ segments of Fig. 12, an expanding quasi-uniform vortex-array
carries a flux-tube array outward with it.  If growing flux-tube-induced stress
on the crust is partly relaxed by ``sudden'' outward crust movements (of 
magnitude $s$) where the stress is strongest (with density preserving backflow 
elsewhere in the stratified crust) the following consequences are expected:
\vspace*{.6cm}

\begin{enumerate}
\item a ``sudden'' permanent increase in $\mu_\perp$, spin-down torque, and 
$|\dot{\Omega}| : { {\Delta \dot{\Omega}}/\dot{\Omega}} \sim s/R
\sim \Delta \theta ~\rm{(strain~relaxation)}~ \lesssim \theta_{max} \sim 
10^{-3}$.
(This is the largest non-transient fractional change in any of the pulsar
observables expected from ``breaking'' the crust.)  A permanent 
glitch-associated
jump in NS spin-down rate of this sign and magnitude ($\sim 3 \times 10 ^{-4}$)
is indeed observed in the larger Crab glitches (Fig. 14)\cite{ref25,ref26,
ref27,ref28}.

\vspace*{.6cm}

\begin{figure*}
\centerline{
\includegraphics*[width=5.5in]{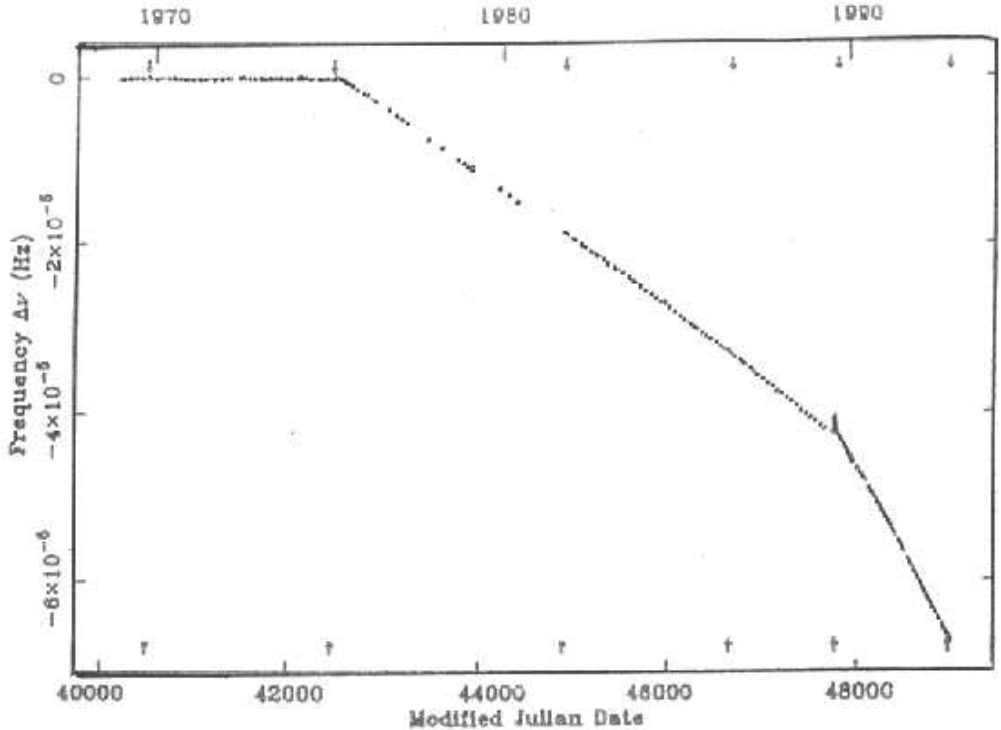}}
\caption{The difference between Crab pulsar periods observed over a 23 yr
	interval and those predicted from extrapolation from measurement
	of $P$, $\dot{P}$, and $\ddot{P}$ at the beginning of that interval.
	These ``sudden'' permanent fractional jumps in spin-down rate
	($\Delta \dot{\Omega}/\dot{\Omega} \sim +5 \times 10^{-4}$) occur
	at glitches ($\Delta \Omega / \Omega \sim 10^{-8}-10^{-7}$)
	but are $10^4$ times greater in magnitude\cite{ref31,ref32}.}
\label{f15}
\end{figure*}

\item a ``sudden'' reduction in shear stress on the crust by the flux-tubes
attached to it from below.  This is matched by an equivalent reduction in 
pull-back on the core's expanding vortex array by the core flux-tube array
attached to it.  The n-vortices therefore ``suddenly'' move out to a new 
equilibrium position where the Magnus force on them is reduced by just this
amount. The high density SF-n sea therefore spins down a bit.
All the (less dense) charged componentes of the NS (crust, core-p and-e)
together with the flux-attached n-vortex-array spin-up much more.
(The total angular momentum of the NS does not change significantly in the brief
time for development of the glitch.)  A new equilibrium is established in which
the charged components (all that can be observed, of course, is $P$ of the 
crust's surface) have been spun up.
For Crab $B$ and $P$, the estimated\cite{ref12} 
${\Delta \Omega}/ \Omega \sim 10^{-4} ({\Delta \dot{\Omega}}/\dot{\Omega})$, consistent with both the 
relatively large Crab glitches of Fig. 14 and also with much smaller Crab 
glitches not shown there\cite{ref28}.
\vspace*{.6cm}
\end{enumerate}


\begin{figure}[tbh]
  \centerline{
    \includegraphics*[height=2.75in]{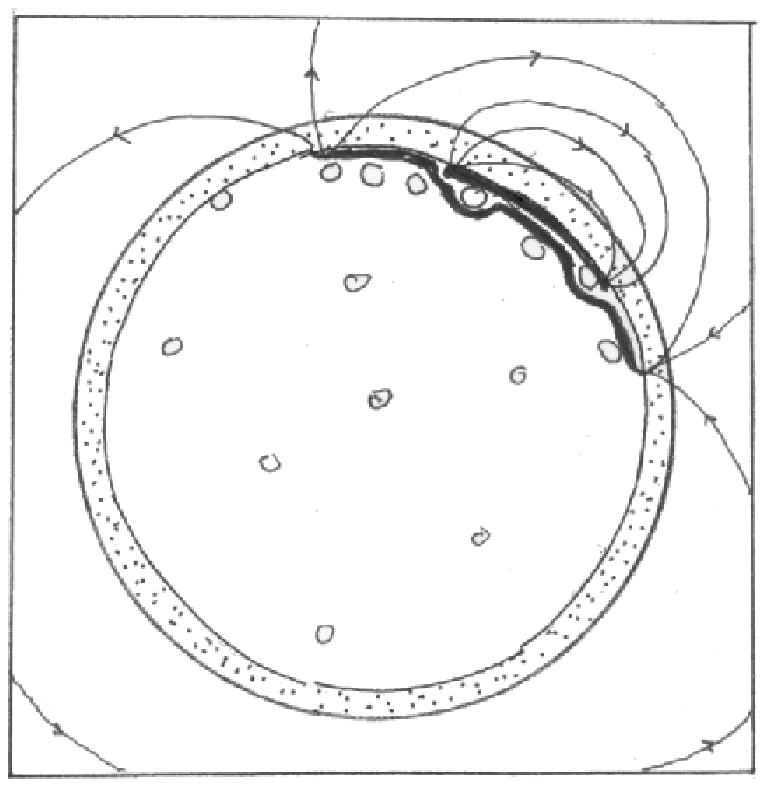}\hfill
    \includegraphics*[height=2.75in]{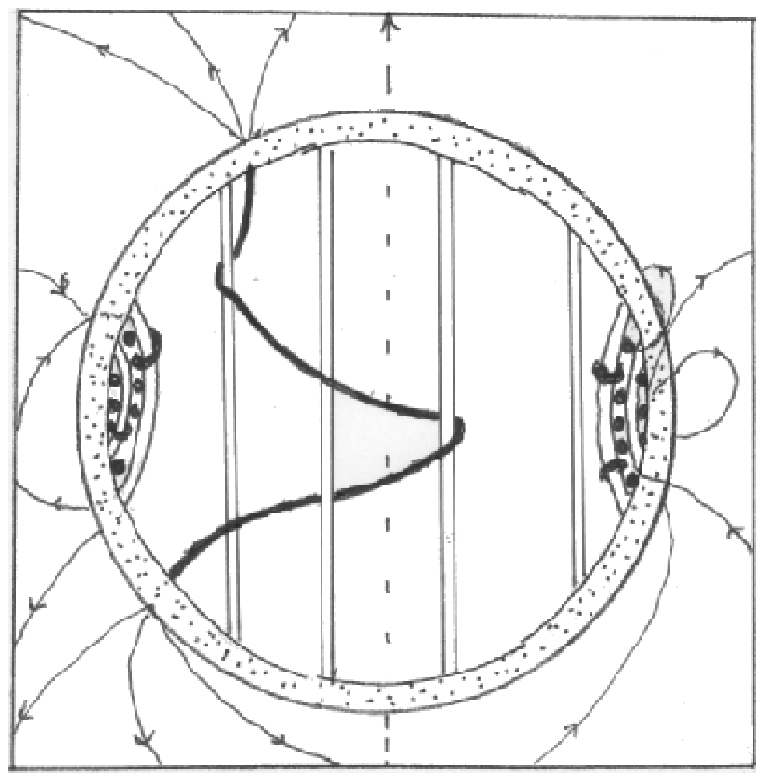}\hfill
   }
  \begin{minipage}{\wcap}
    \caption{The configuration (top view) of Fig. 11 after further spin-down.
	Flux-tubes are piling up in an equatorial annulus at the core-crust
	interface.  The blocked flux-tubes, in turn, block short segments
	of vortex lines which forced them into this annulus.}
    \vspace{.3cm}
  \end{minipage} \hfill
  \vspace{.6cm}
  \begin{minipage}[tr]{\wcap}
    \caption{A side view of the representation of the Fig. 15 configuration
	with the addition of one flux-tube, which the expanding vortex-array
	has not yet forced out to a radius $\sim R$.}
    \vspace{1cm}
  \end{minipage}
\end{figure}


\item  {\emph{Giant Vela-like (V) glitches.}} The second (V)-family of
glitches differs from that of Crab-like ones (C) in several ways.

\begin{enumerate}
\vspace*{.6cm}

\item ${(\Delta \Omega}/ \Omega)_V \sim 10^2 \times (\Delta \Omega/
 \Omega)_C$.
\vspace*{.6cm}

\item V-glitches develop their $\Delta \Omega$ in less than $10^2$ sec.:
the $\Delta \Omega$ of a V-glitch is already decreasing in magnitude when first
resolved\cite{ref26}, while C-glitches are still rising toward their full 
$\Delta \Omega$ for almost $10^5$ sec\cite{ref34,ref35}. 
\vspace*{.6cm}

\item V-glitches are observed in pulsars (mainly, but not always) in Fig. 12 
along $(C \rightarrow D)$ while C-glitches are observed all along 
 $(A \rightarrow C \rightarrow D)$.
\vspace*{.6cm}

\item The C-glitch proportionality between $\Delta \dot \Omega / \dot{\Omega}$ 
and $\Delta \Omega / \Omega$ 
would greatly overestimate (${\Delta \dot{\Omega}} / \dot{\Omega}$) for 
V-glitches.  
\vspace*{.4cm}

\begin{figure*}[h*]
  \centerline{
    \includegraphics[height=2.75in]{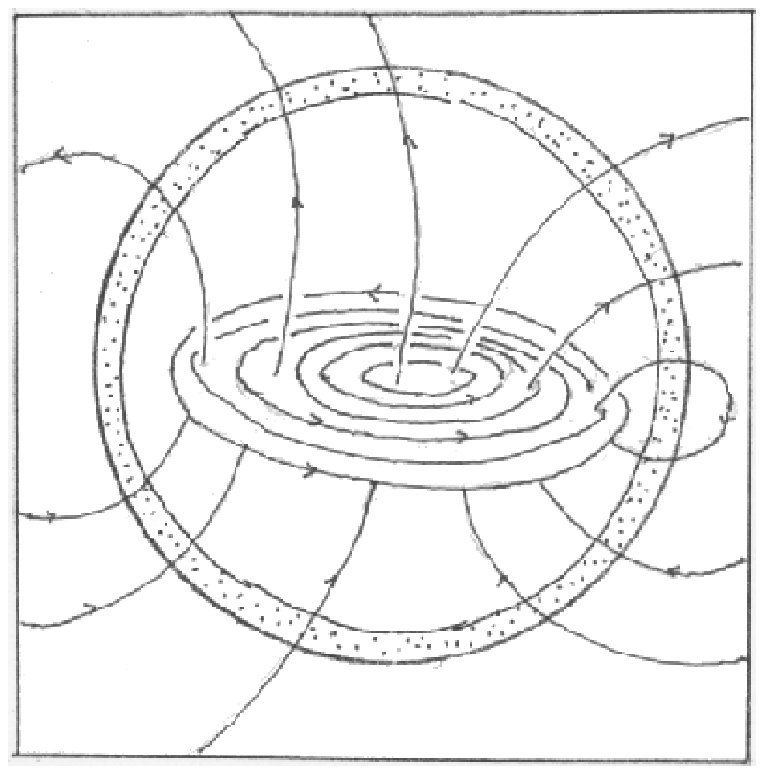}\hfill
    \includegraphics[height=2.75in]{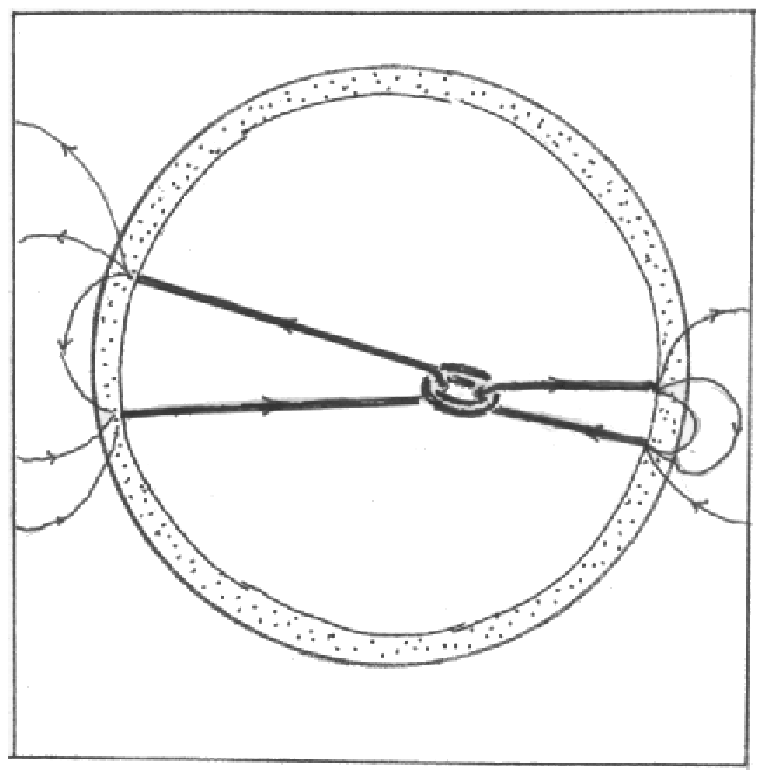}\hfill
   }
  \begin{minipage}{\wcap}
    \caption{A schematic representation of a young NS's magnetic field just
	before the NS cools to the transition temperature for proton
	superconductivity.  Some shearing stress preventing an even more
	stabilized configuration is probably borne by the NS crust which
	solidified much earlier.}
    \vspace{.6cm}
  \end{minipage} \hfill
  \vspace{.6cm}
  \begin{minipage}[tr]{\wcap}
    \caption{A representation of the Fig. 17 magnetic field after core flux-tube
	formation and relaxation to a new quasi-equilibrium.  The initially
	increased stress in the crust (cf. Fig. 1) is assumed to exceed the 
	crust's shear-stress yield strength.  A later formation of SF-n
	vortex-lines would halt such relaxation.}
    \vspace{.6cm}
  \end{minipage}
\end{figure*}

The existence of a second glitch family, with 
V-properties, is expected from a second effect of vortex-driven flux-tube 
movement in a NS core.  If there were no very dense, comoving, flux-tube 
environment around them, outward moving core-vortices could smoothly shorten
and then disappear entirely as they reached the core's surface at its 
spin-equator. (We ignore crustal SF-n here.)  
However, the strongly conducting crust there resists entry of the
flux-tubes which the vortices also bring with them to  the crust's base.
This causes a pile-up of pushed flux-tubes into
 a small equatorial annulus
(Figs. 15 and 16)
which delays the final vortex-line disappearance. The vortex movement in which
they vanish occurs either in vortex-line flux-tube cut-through events, or, 
more likely, in a sudden breaking of the crust which has been overstressed by 
the increasing shear-stress on it from the growing annulus.
Giant V-glitches were proposed as such events\cite{ref12,ref8}, allowing a 
``sudden'' reduction of part of this otherwise growing annulus of excess angular
momentum and also some of the magnetic flux trapped within it.
These would not begin until enough vortex-lines, initially distributed almost
uniformly throughout the core, have piled up in the annulus 
for the flux-tubes they bring with them to supply the needed shear stress.
Estimates of V-glitch $\Delta \dot{\Omega} / \dot{\Omega}$ magnitudes are
less reliable than those for C-glitch ones.
A very rough one, based upon plausible guesses and an assumed $\Omega/R$ 
about the same as those in the larger C-glitches, suggest V-glitch 
repetition rates and magnitudes not unsimilar to observed ones\cite{ref8,ref12}.
\end{enumerate}
\end{list}

\begin{center}
\begin{figure}[ht*]
\centerline{
\includegraphics[width=5.5in]{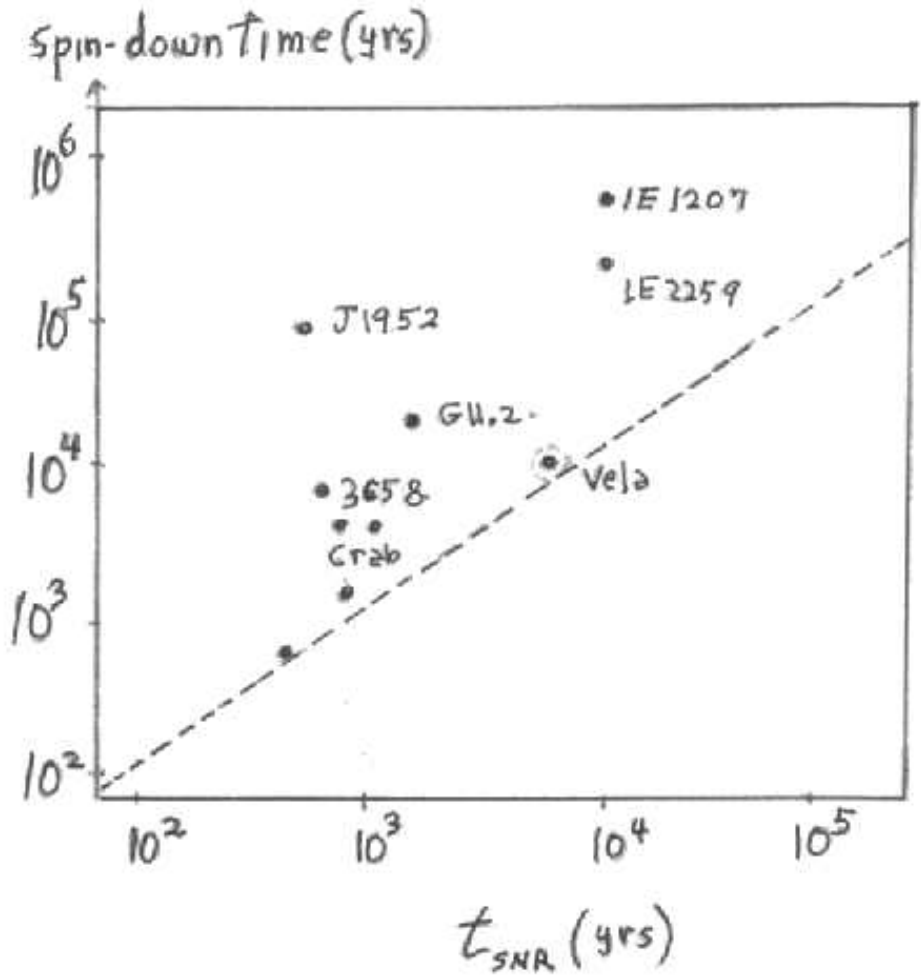}}
\caption{Observed spin-down times for pulsars ($P/2\dot{P}$) vs the time since
	birth of these same pulsars as inferred by the ages of the supernova
	remnants in which they are still embedded ($t_{SNR}$)\cite{
	ref8, ref33}. 
}
\label{f20}
\end{figure}
\end{center}

\section{In the beginning}

The proposed spin-down biography of a NS surface $B$ presented in Sects. 3,4,
and 5 began at $(A)$ (or perhaps $A'$) in Fig. 12 when that typical NS is 
expected to be about $10^3$ yrs old.
Before that its crust had solidified (age $\sim$ a minute), 
its core protons had become superconducting ($\sim 1$  yr?), 
and core neutrons became superfluid ($\sim 10^3$ yrs?).  
If so, there would be a nearly $10^3$ year interval between
formation of the NS core's magnetic flux-tube array and control of that array's
movement by that of a SF-n vortex array.
During that interval an early magneto-hydrodynamic equilibrium involving 
poloidal and toroidal fields, and some crustal shear stress (Fig. 17)
would be upset by the dramatically altered $B$-field stresses after
flux-tube formation\cite{ref8}.  
The subsequent jump in shearing stress on the crust surface $B$
change.  
The recent reconsideration of drag on moving flux-tubes\cite{ref30} suggests
the core flux-tube adjustment can take $\sim 10^3 $ yrs. 
For many NSs, depending on historical details of their $B$ structure, dipole
moments should become much smaller (Fig. 18). 
Their post-partem values and subsequent expected drops in their sizes have been
estimated and proposed \cite{ref8} as the reason many young pulsars have
spin-down ages ($P/2\dot{P}$) up to $10^2$ times greater than their true ages
(Fig. 19).

\section{Acknowledgements}

I am happy to thank E.V. Gotthelf, J.P. Halpern, P. Jones, J. Sauls, J.Trumper,
and colleagues at the Institute of Astronomy (Cambridge) for helpful 
discussions.

\end{document}